\documentclass[sigconf]{acmart}

\AtBeginDocument{%
  \providecommand\BibTeX{{%
    \normalfont B\kern-0.5em{\scshape i\kern-0.25em b}\kern-0.8em\TeX}}}

\copyrightyear{2023} 
\acmYear{2023} 
\setcopyright{acmlicensed}\acmConference[WWW '23 Companion]{Companion Proceedings of the ACM Web Conference 2023}{April 30-May 4, 2023}{Austin, TX, USA}
\acmBooktitle{Companion Proceedings of the ACM Web Conference 2023 (WWW '23 Companion), April 30-May 4, 2023, Austin, TX, USA}
\acmPrice{15.00}
\acmDOI{10.1145/3543873.3584627}
\acmISBN{978-1-4503-9419-2/23/04}

\newcommand{\xy}[1]{\textcolor[rgb]{0,0,0}{#1}}

\begin{document}

\title{MAKE: Vision-Language Pre-training based Product Retrieval in Taobao Search}

\author{Xiaoyang Zheng, Zilong Wang, Sen Li}
\affiliation{%
  \institution{Alibaba Group}
  \city{Hangzhou}
  \country{China}
    \institution{\\ \{zhengxiaoyang.zxy,huanshi.wzl,\\lisen.lisen\}@alibaba-inc.com
    }
}

\author{Ke Xu}
\authornote{Corresponding author}
\affiliation{
  \institution{City University of Hong Kong}
  \city{Hong Kong}
  \country{China}
  \institution{\\kkangwing@gmail.com}
}
    
\author{Tao Zhuang, Qingwen Liu, Xiaoyi Zeng}
\affiliation{%
 \institution{Alibaba Group}
 \city{Hangzhou}
 \country{China}
   \institution{\\ 
   \{zhuangtao.zt,xiangsheng.lqw,\\yuanhan\}@alibaba-inc.com
   }
}

\renewcommand{\shortauthors}{Zheng, et al.}

\begin{abstract}
Taobao Search consists of two phases: the retrieval phase and the ranking phase. Given a user query, the retrieval phase returns a subset of candidate products for the following ranking phase. 
Recently, the paradigm of pre-training and fine-tuning has shown its potential in incorporating visual clues into retrieval tasks. In this paper, we focus on solving the problem of text-to-multimodal retrieval in Taobao Search.
We consider that users' attention on titles or images varies on products. Hence, we propose a novel \textbf{M}odal \textbf{A}daptation module for cross-modal fusion, which helps assigns appropriate weights on texts and images across products. 
Furthermore, in e-commerce search, user queries tend to be brief and thus lead to significant semantic imbalance between user queries and product titles. Therefore, we design a separate text encoder and a \textbf{K}eyword \textbf{E}nhancement mechanism to enrich the query representations and improve text-to-multimodal matching.
To this end, we present a novel vision-language (V+L) pre-training methods to exploit the multimodal information of (user query, product title, product image). 
Extensive experiments demonstrate that our retrieval-specific pre-training model (referred to as \textbf{MAKE}) outperforms existing V+L pre-training methods on the text-to-multimodal retrieval task.
\textbf{MAKE} has been deployed online and brings major improvements on the retrieval system of Taobao Search.
\end{abstract}




\maketitle

\section{Introduction}
Online shopping has become popular in our daily lives. Hundreds of millions of users visit e-commerce platforms (such as Amazon, eBay, Taobao, and JD) every day. The product search service, which displays relevant products based on user queries, is of great importance for user experience and transaction efficiency. 
Taobao Search consists of two phases: the retrieval phase and the ranking phase. The retrieval phase aims to select a candidate set (tens of thousands) from a large pool of products (in billion level), while the ranking phase determines the displaying order. Hence, the retrieval phase plays an important role in the quality of search results. In Taobao\footnote{https://www.taobao.com/}, a product post is composed of a title and several images, while user queries are plain texts. Therefore, the retrieval phase is formulated as a problem of text-to-multimodal matching.

There are many works \cite{li2021embedding,nigam2019semantic,xiao2019weakly,chang2021extreme,zobel2006inverted,robertson2009probabilistic} proposed for the product retrieval task, which fall into two categories:
lexical matching approaches and embedding-based learning approaches. Lexical matching approaches \cite{zobel2006inverted,robertson2009probabilistic} typically build the inverted indexes for products and conduct exact matching between inverted indexes and user queries. 
Embedding learning approaches \cite{li2021embedding,nigam2019semantic,xiao2019weakly,chang2021extreme} learn semantic representations (i.e., embeddings) of queries and products, and then retrieve products by measuring the similarity between the query and product embeddings.


\begin{table}[htbp]
    \centering
    \footnotesize
    \begin{tabular}{p{0.14\linewidth} | p{0.79\linewidth}}
    \toprule
        Query & Title \\\midrule
        White shirt & White chiffon shirt, women's long-sleeved top, 2020 spring and autumn new western style professional wear, light mature temperament \\ 
    \bottomrule
    \end{tabular}
    \caption{An example of query-title pair collected from online logs. There exists a significant imbalance between user queries and product titles.
    }
    \label{tab:query-title}
\end{table}


Recently, the success of transformer \cite{vaswani2017attention} structure and vision-language representation learning \cite{zhang2021vinvl,qi2020imagebert,li2021align} motivates people to study pre-training on e-commerce tasks \cite{gao2020fashionbert,zhuge2021kaleido,yu2022commercemm}. These models, composed of a text encoder and an image encoder based on transformers, are pre-trained on text-image pairs and fine-tuned on image captioning, category recognition, text-to-image retrieval, etc.
Intuitively, to solve the text-to-multimodal retrieval task in Taobao Search, we exploit the text encoder on user queries and product titles, while applying the image encoder on product images. The representations of user queries and products are then used in an embedding retrieval framework. However, we observe sub-optimal performance due to the following two key problems.
%
\textbf{First}, existing methods neglect the fact that in e-commerce search, users' attention to titles or images varies on products. For example, users pay more attention to images of clothes. Whereas on electronic products, users care more about key properties described in titles, such as memory size. 
\textbf{Second}, sellers often apply search engine optimization (SEO) techniques, in order to improve the matching probabilities and ranking of their products. As shown in Table \ref{tab:query-title}, user queries are usually short and brief, while product titles tend to be long and concrete. The semantic imbalance between user queries and products is a big challenge for the retrieval task. %

%
%
To handle the \textbf{first} problem, we propose a \textbf{M}odal \textbf{A}daptation module to perform cross-modal fusion by introducing user queries as contextual information and to assign reasonable attention to product titles and images.
To address the \textbf{second} issue, we design an independent text encoder to process user queries. We further design a \textbf{K}eyword \textbf{E}nhancement mechanism by jointly optimizing similar positive samples for user queries, in order to enrich the semantic information and learn better user query embeddings. 

%

%

To summarize, our main contributions are as follows: 
\begin{itemize}
    \item We propose a novel vision-language pre-training method (referred as \textbf{MAKE}) tailored for the text-to-multimodal retrieval task in e-commerce search. Trained on a large-scale (\textit{query}, \textit{title}, \textit{image}) triplet dataset from online logs of Taobao Search, \textbf{MAKE} is capable of effective and efficient text-to-multimodal retrieval. 
    \item We propose a \textbf{M}odal \textbf{A}daptation module to learn appropriate attentions on product titles and images by introducing user queries as the context. The module leads to stronger representation power of product embeddings.
    \item We propose a \textbf{K}eyword \textbf{E}nhancement mechanism to enhance the query embeddings by jointly training similar user queries. The mechanism significantly alleviates the semantic imbalance between user queries and product titles. 
    \item 
    Extensive experiments on offline datasets and online A/B tests demonstrate that \textbf{MAKE} outperforms existing V+L pre-training methods on the e-commerce text-to-multimodal retrieval task. Our method has been deployed on Taobao Search and served hundreds of millions of users every day.
\end{itemize}

\section{The Proposed Approach}
\begin{figure}[htbp]
    \centering
    \includegraphics[width=0.45\textwidth]{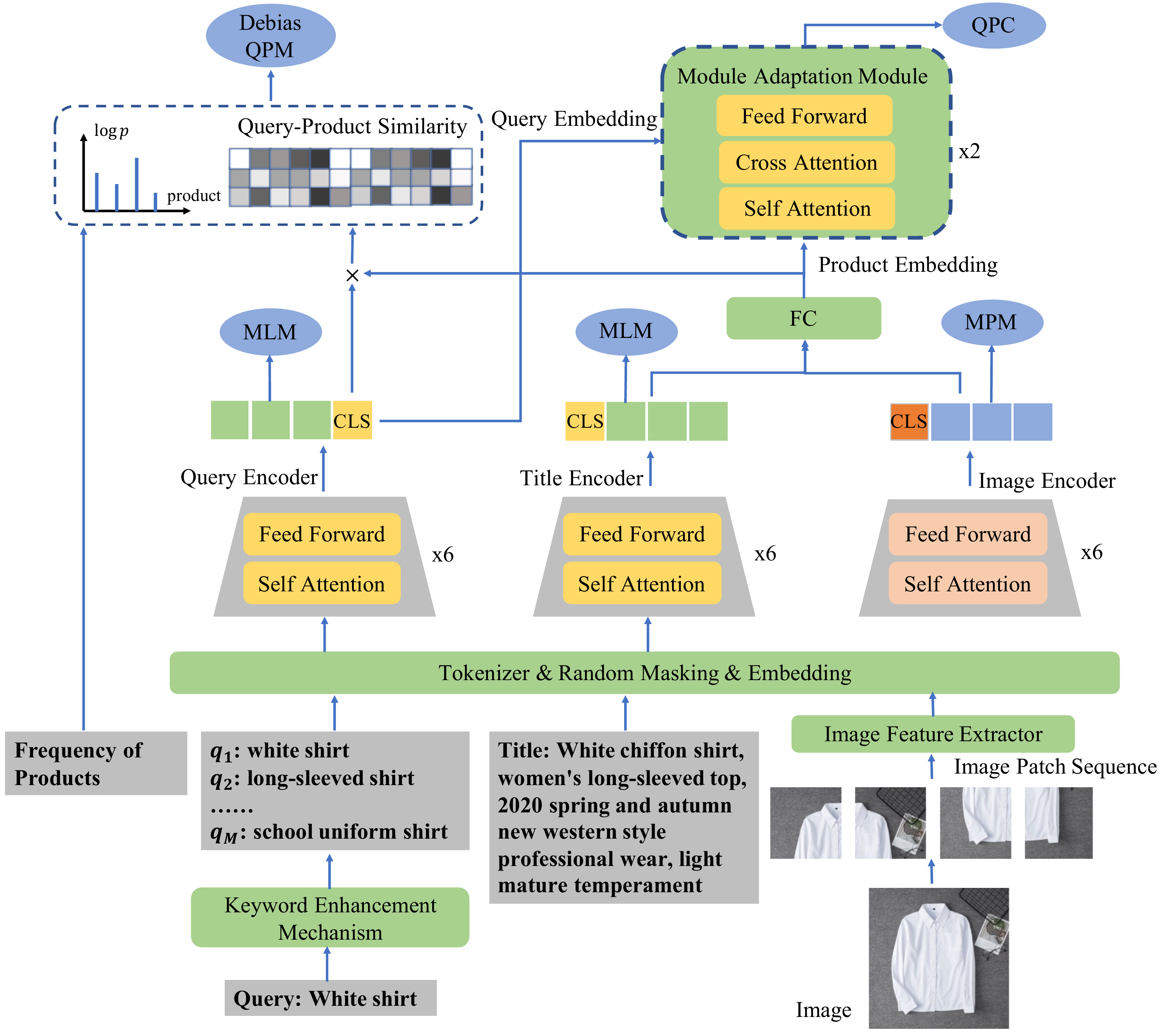}
    \caption{Overview of our pre-trained model MAKE.}
    \label{fig:network}
    \Description{Network Architecture}
\end{figure}

\subsection{Vision-Language Pre-training Model}
\label{sec:pretrain}
\paragraph{\textbf{Model Structure.}} Different from existing V+L methods with a text encoder and an image encoder, our pre-training model has a three-tower structure, with a query encoder, a title encoder and an image encoder, as shown in Figure \ref{fig:network}. 
Each of them consists of 6 transformers (a multi-head self-attention layer and a feed-forward layer) blocks~\cite{vaswani2017attention}. The two text encoders are initialized with the Structbert \cite{wang2019structbert} model (pre-trained on Chinese e-commerce corpus), and the image encoder is initialized with the ImageBert \cite{qi2020imagebert} model (pre-trained on Chinese corpus and corresponding images \cite{qiu2021easytransfer}). The embeddings of products are the combination of outputs from the title encoder and the image encoder.

\paragraph{\textbf{Model Inputs.}} The text modal (user queries and product titles) is pre-processed in the same way as BERT \cite{devlin2019bert}. We adopt the Chinese vocabulary provided by EasyTransfer \cite{qiu2021easytransfer}. For the product images, we preprocess them as 4*4 patches and apply ResNet \cite{he2016deep} as a backbone network to extract image sequences of 2048-D features. Segmentation marks \textit{Q}, \textit{T} and \textit{I} are used to distinguish token sequences of user queries, product titles, and images.

\paragraph{\textbf{Self-Supervised Pre-training Objective.}} Pre-training models with self-supervised task \cite{devlin2019bert,yu2022commercemm} are proved to be effective on many downstream tasks. Following FashionBert \cite{gao2020fashionbert}, we apply two self-supervised tasks: Masked Language Modeling (MLM) and Mask Patch Modeling (MPM). Refer to the paper for detailed information.

\paragraph{\textbf{Query Encoder.}} To start with, we pre-train ALIGN \cite{li2021align} on image-text pairs from Taobao. We apply the same text encoder to user queries and product titles. Nevertheless, we observe sub-optimal performance on the query-to-product retrieval task. We find out that in e-commerce search, it is common for sellers to apply search engine optimization (SEO) techniques for improving the rankings of their products.
As a result, the product titles usually consist of many keywords that are grammatically meaningless and contain grammatical errors. On the contrary, users tend to type in short terms to the search engine. Hence, there exist non-trivial imbalances between queries and titles. Therefore, we need a separate text encoder for user queries. Besides, for V+L pre-training models \cite{li2021align,qi2020imagebert}, Image-Text Matching (ITM) is widely used to improve the performance on the downstream retrieval task. Similar to ITM, we adopt a Query-Product Matching (QPM) loss with in-batch negative sampling to optimize embeddings of queries and products. 
\begin{equation}
    \mathcal{L}_{QPM}=-\frac{1}{N}\sum_i^N\log\frac{\exp(\boldsymbol{u}^T\boldsymbol{v}/\tau)}{\sum_{j=1}^N\exp(\boldsymbol{u}^T\boldsymbol{v_j}/\tau)}, \label{eq:qpm}
\end{equation}
where $\boldsymbol{u}$ and $\boldsymbol{v}$ are normalized embeddings of queries and products. $N$ is the batch size and $\tau$ is the temperature parameter.
\subsection{Modal Adaptation Module}
\label{sec:ma}
In e-commerce, it is obvious that the importance of titles and images varies across different products. 
For instance, on clothes, users pay more attention to images. Whereas on electronic products, users care more about key properties described in titles, such as memory size. Hence, two modalities should be fused with proper weights for different products. However, we observe that with separated text/image encoders, the pre-training model focuses evenly on text/image modals. We believe that the lack of cross-modal fusion prevents the network from learning better representations.

Although Yu \etal\cite{yu2022commercemm} design a multimodal fusion encoder on top of the text/image encoder, they ignore user intentions. Therefore, by introducing user queries as contextual information, we propose a novel \textbf{M}odal \textbf{A}daptation module in order to conduct modal fusion and optimize the overall representations, as shown in Figure \ref{fig:network}. The module contains two layers of a sub-module, composed of a self-attention layer, a cross-attention layer and a feed-forward layer, which takes outputs of the query encoder, title encoder, and image encoder as inputs.
For the self-attention layer, inputs of \textit{KV} are outputs of the title encoder and the image encoder, while for the cross-attention layer, inputs of \textit{Q} are outputs of the query encoder. 
With the Modal Adaptation module, embeddings of products not only contain two-modal (title+image) information of diverse weights within products but also consider the influence of user queries. 

We also design a Query-Product Classification (QPC) loss for the \textbf{MA} module. Different from optimizing similarity among separated embeddings in the QPM loss, the QPC loss refines the joint representation of query-product pairs. 
The [CLS] outputs (representations of the whole sequence) of the \textbf{MA} module, followed by a fully-connected layer and a sigmoid function, are used in a two-class classification task.
We construct negative query-product pairs by choosing the maximum similarity among mini-batch negative samples. The similarity is pre-computed in QPM loss, and thus brings little computational cost. The QPC loss is presented as:
\begin{align}
\mathcal{L}_{\mathrm{QPC}}&=\mathbb{E}_{(Q, T, I) \sim D} \mathrm{H}(\boldsymbol{p}_{\mathrm{QPC}}(Q, T, I), \boldsymbol{y}_{\mathrm{QPC}}),
\end{align}
where $\boldsymbol{p}_{QPC}$ is the probability of classification, $\boldsymbol{y}_{\mathrm{QPC}}$ is a 0-1 ground-truth label and $\mathrm{H}$ is the cross-entropy loss.

\subsection{Keyword Enhancement Mechanism}
\label{sec:ke}
As mentioned above, the QPM (Eq. \ref{eq:qpm}) loss is associated with in-batch negative sampling (IBNS) adopted in ALIGN \cite{li2021align}, which brings another significant issue to our model. Different from limited academic tasks, multiple user queries may be relevant to the same product. With the IBNS mechanism, those similar user queries are mistakenly treated as negative samples and thus compromise query embeddings. Therefore, to solve the false-negative issue of the IBNS mechanism, we propose a Keyword Enhancement mechanism to replace the IBNS mechanism. The proposed mechanism aims at improving representation learning of user queries by jointly optimizing queries related to the same product. 

Instead of \textit{query-product} pairs, a product with several related queries ($\textit{product}$, $\textit{query}_1$, $\ldots$, $\textit{query}_M$) collected from Taobao Search logs are grouped as one training sample. $M$ is the number of enhanced queries and is set to 5 in this paper. 
In addition, we design a new QPM loss based on circle loss \cite{sun2020circle} and \textbf{KE} mechanism:
\begin{equation}
    \mathcal{L}_{QPM}^{KE} = \log(1+\sum_{j=1}^N\exp(\gamma\exp(s_{neg}^j+\theta))\sum_{m=1}^M\exp(-\gamma s_{pos}^m)),
\end{equation}
where $s(\cdot)=\boldsymbol{u}^T \boldsymbol{v}-\log \boldsymbol{p}$ measures the inner-product similarity between embeddings of queries and products. Following sampled-softmax \cite{jean2014using}, the $-\log\boldsymbol{p}$ term is the expected frequency of products, with which we prevent the model from focusing too much on popular products. $N$ is the batch size.
$\gamma$ is the scaling factor. The hyper-parameter $\theta$ constrains the lower bound of the similarity difference between positive pairs and negative pairs. With the Keyword Enhancement mechanism, we solve the false-negative issue and narrow the distance between embeddings of similar queries.

Finally, the pre-trained model is optimized as below:
\begin{equation}
\mathcal{L}=\mathcal{L}^Q_{MLM}+\mathcal{L}^T_{MLM}+\mathcal{L}^{I}_{MPM}+\mathcal{L}_{QPC}+\mathcal{L}_{QPM}^{KE}.
\end{equation}

\section{Experiments}
\subsection{Datasets, Implementations, and Metrics}

\textbf{Large-scale Industrial Dataset.} We collect online clicking logs with user queries, product titles and images from Taobao Search. The training set contains samples of billion-level, and we randomly choose 1.5 million search logs as the evaluation set. 



\textbf{Model Implementation.} 
The pre-training model is composed of three encoders with 6 layers of transformers \cite{devlin2019bert}. Each layer has $768$ hidden units and 12 self-attention heads. We pre-train the model for 10 epochs with a batch size of $1280$ on $50$ NVIDIA P100 GPUs. We applied an Adam optimizer with $\beta_1=0.9$ and $\beta_2=0.98$. The learning rate is warmed-up to $1e^{-4}$ in the first $2000$ iterations and decays to $0$ following a linear schedule.

\textbf{Online Serving.} We predict embeddings of all products in Taobao with the title encoder and the image encoder. Then we adopt Proxima \cite{proxima}, an ANN (approximate nearest neighbor) framework, to build indexes of product embeddings with HC (hierarchical clustering) algorithm. Once receiving a user request, the online query encoder predicts the user query and returns the embedding. The query embedding is used to retrieve top-K relevant products from the ANN index. The model is updated on weekly basis.

\textbf{Offline Evaluation Metrics.} The retrieval set is denoted as $R=\{p_1,\ldots,p_K\}$. The clicked products from the evaluation set is denoted as the target set $T$.
We use the metric $P_{rel}$ and $P_{cate}$, which measures the rate of relevance on the retrieval set $R$, according to a well-trained relevance model \cite{DBLP:conf/www/YaoTCYXD021} (the AUC on human-labeled data is $0.92$). The first one focus on the overall relevance, while the second one compares the category predicted on user queries to the category of retrieved products. 
\begin{equation}
\small
    P_{rel}=\frac{1}{NN_R}\sum_{i=1}^N\sum_{j=1}^{N_R} f(q_i,p_{i,j}), \quad P_{cate}=\frac{1}{NN_R}\sum_{i=1}^N\sum_{j=1}^{N_R} \mathbb{I}(f_c(q_i)=c_{i,j}),
\end{equation}
where $f(\cdot,\cdot) \in[0, 1]$ denotes the prediction of the relevance model, $f_c(\cdot)$ returns category based on queries. $N$ is the size of the evaluation dataset and $N_{R_{i}}$ is the size of retrieval set $R_i$. $c_{i,j}$ is the category of the retrieved product $p_{i,j}$. 
We also apply a Recall@K metric to evaluate the retrieval performance, computed as:
\begin{equation}
\text{Recall@K}=\frac{1}{N}\sum_{i=1}^{N}\mathbb{I}(\exists t | t\in R_{i,K} \wedge t\in T_i),
\end{equation}
where $\mathbb{I}(\cdot)$ is an indicator function. 


\textbf{Online Evaluation Metrics.} We use the number of transactions (denoted as \#Trans) and GMV (Gross Merchandise Volume, \xy{total value of sales}) as online evaluation metrics.
For users with few recorded consuming behaviors, these two metrics are denoted as $\text{\#Trans}_n$ and $\text{GMV}_n$, respectively.

\subsection{Offline Experimental Results}
\begin{figure}[htbp]
    \centering
    \includegraphics[width=0.43\textwidth]{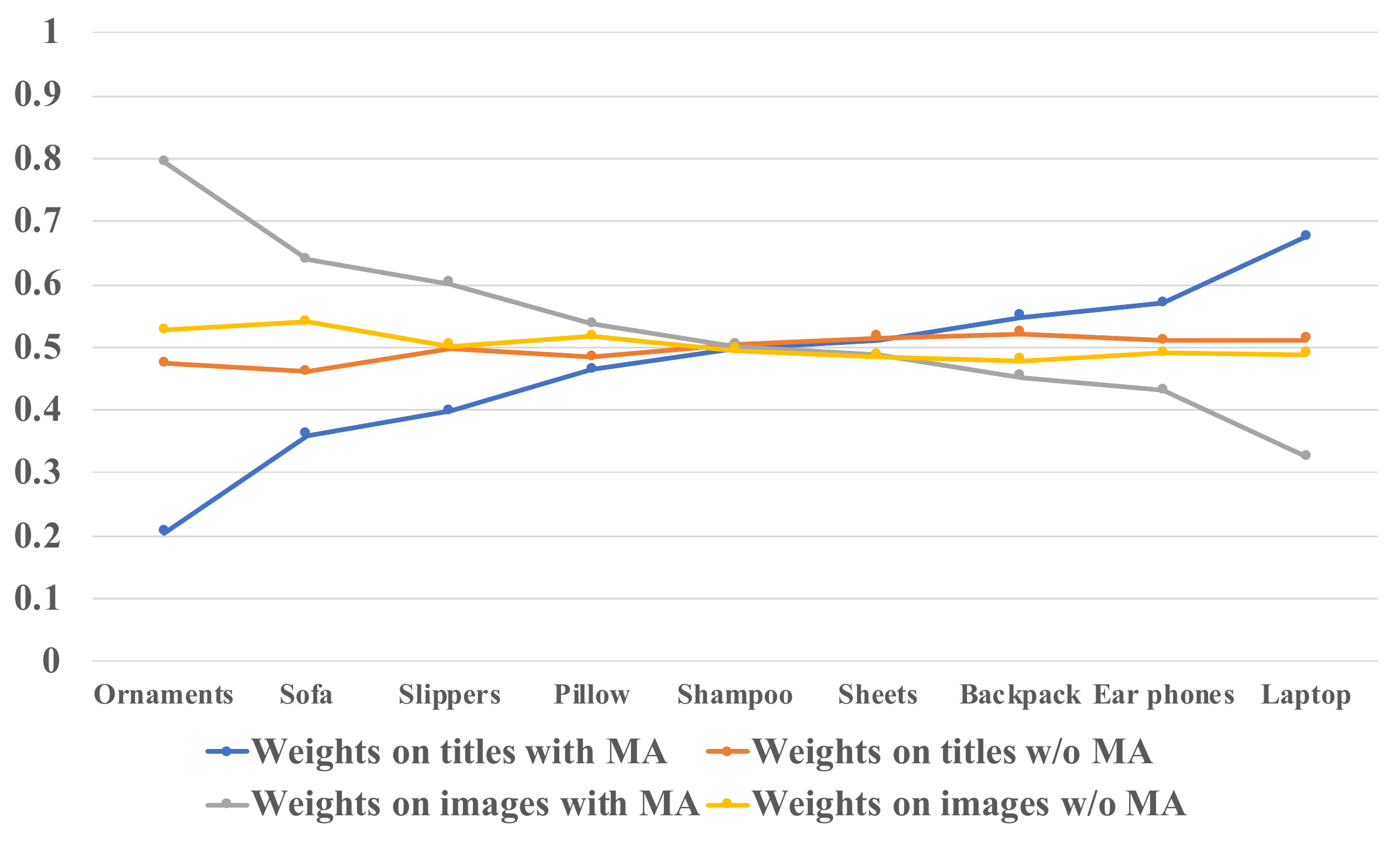}
    \caption{Attention weights on titles and images across different product categories. MAKE vs. MAKE w/o MA.
    \Description{Attentions weights on titles and images}
    }
    \label{fig:weight}
    \Description{Attention weights on two modals.}
\end{figure}


\subsubsection{Comparison with Baseline Methods}
We adopt CLIP \cite{radford2021learning}, FashionBERT \cite{gao2020fashionbert} and CommerceMM \cite{yu2022commercemm} as strong baseline pre-training models. The latter two are proposed to solve downstream tasks related to the e-commerce scenario. FashionBERT \cite{gao2020fashionbert} consists of one encoder, while CLIP \cite{radford2021learning} and CommerceMM \cite{yu2022commercemm} have a two-tower structure. 
All baseline methods are pretrained on the same training dataset. 
As shown in Table \ref{tab:ablation}, our proposed method \textbf{MAKE} outperforms all baseline methods on the text-to-multimodal retrieval task of Taobao Search.
\subsubsection{Modal Adaptation Module. } The \textbf{MA} module, associated with the Query-Product Classification (QPC) task, is proposed to conduct modal fusion and learn appropriate attention on text/image modals across different products.
%
The comparison from \textbf{MAKE w/o MA} to \textbf{MAKE} reveals that the \textbf{MA} module significantly improves the relevance and the recall hitrate by 1.88\% and 4.59\%. 
To further evaluate the effect of \textbf{MA} module, we collect attention weights on titles/images across different product categories from \textbf{MAKE} and \textbf{MAKE w/o MA}, as shown in Figure \ref{fig:weight}.
After introducing the \textbf{MA} module, for vision dominant categories, the network pays more attention to images ($79.4\%$ for ornaments), while for text dominant categories, the network focuses more on titles ($67.5\%$ for laptops). With the \textbf{MA} module, the model assigns proper attention weights on modals of text and image.
\begin{table}[htbp]
    \centering
    \begin{tabular}{lccc}
    \hline
        Methods & $P_{rel}\uparrow$ & $P_{cate}\uparrow$ & Recall@$K\uparrow$ \\\hline
        FashionBert \cite{gao2020fashionbert} & 0.8385 & 0.8190 & 0.3867 \\\hline
        CLIP \cite{radford2021learning} & 0.8648 & 0.8423 & 0.4675\\\hline
        CommerceMM \cite{yu2022commercemm} & 0.8710 & 0.8653 & 0.4937 \\\hline
        MAKE & \textbf{0.9014} & \textbf{0.9295} & \textbf{0.6088} \\\hline
        MAKE w/o MA & 0.8826 & 0.8910 & 0.5629 \\\hline
        MAKE w/o KE & 0.8922 & 0.8803 & 0.5781 \\\hline
    \end{tabular}
    \caption{Offline experimental results and ablation studies.
    }
    \label{tab:ablation}
\end{table}

\subsubsection{Keyword Enhancement Module}
%
The \textbf{KE} mechanism is a modified negative sampling mechanism with a QPM loss.
The \textbf{KE} mechanism enforces the model to jointly optimize similar queries and avoid false-negative sampling of the popular IBNS mechanism. By comparing \textbf{MAKE} to \textbf{MAKE w/o KE}, the \textbf{KE} mechanism effectively strengthens the query representations by reducing the distance from similar queries to the same product. The \textbf{KE} mechanism also helps alleviate the semantic imbalance between user queries and product titles.

\subsection{Online A/B Tests}

We deploy our pre-training method \textbf{MAKE} on Taobao Search and provide relevant candidates to the prior three-channel retrieval system, including embedding-based learning, collaborative filtering, and inverted-index matching.
As shown in Table \ref{tab:AB}, our method outperforms the prior retrieval system 
by improving the overall relevance ($+2.20\%$) of product candidates. We also report 14-day average online improvements of \textbf{MAKE} on GMV and \#Trans. As shown in Table \ref{tab:AB}, our proposed method improves GMV and \#Trans by $0.79\%$ and $0.37\%$, respectively. Considering the large number of transactions in Taobao Search, \textbf{MAKE} facilitates hundreds of thousands of transactions per day. Besides, \textbf{MAKE} obtains more performance gains ($2.01\%$ on GMV and $1.58\%$ on \#Trans) on inactive users and new users, including tens of millions of users per day. Compared to the significant gains, the additional computational cost is negligible (\textbf{2 ms}). These results demonstrate that \textbf{MAKE} significantly improves the overall efficiency of Taobao Search.

%
%
\begin{table}[htbp]
    \centering
    \small
    \begin{tabular}{l|cccccc}
    \hline
        Methods & GMV & \#Trans & $\text{GMV}_n$ & $\text{\#Trans}_n$ & $P_{rel}$ & Time Cost \\\hline
        MAKE & +0.79\% & +0.37\% & +2.01\% & +1.58\% & +2.20\% & +2 ms \\\hline 
    \end{tabular}
    \caption{Online A/B tests of MAKE. 
        }
    \label{tab:AB}
\end{table}


\section{Conclusion}
In this paper, we propose a novel vision-language pre-training method (\textbf{MAKE}) with a three-encoder structure tailored for the text-to-multimodal retrieval task of Taobao Search. We propose a Modal Adaptation module to perform cross-modal fusion and learn effective product representations. We further design a Keyword Enhancement mechanism to solve the semantic imbalance and false-negative sampling issue to improve query representations.
Offline ablation study and online A/B tests, verify the effectiveness of our method \textbf{MAKE}. We have deployed \textbf{MAKE} online to serve hundreds of millions of users every day, which greatly improves online transaction efficiency in e-commerce. 


\bibliographystyle{ACM-Reference-Format}
\bibliography{sample-base}








\end{document}